# SubGraD- An Approach for Subgraph Detection

Akshara Pande[1], Vivekanand Pant[2,] S. Nigam[1]

[1] Human Resource Development Group,
CSIR Complex, Library Avenue, Pusa, New Delhi, India.
pandeakshara@gmail.com, snigam@csirhrdg.res.in
[2] IBM, A-26, Sector- 62, NOIDA, Gautam Buddha Nagar,
Noida - 201307, India.
vickypantmnr@gmail.com

## ABSTRACT

*A new approach of graph matching is introduced in this paper, which efficiently solves the problem of graph isomorphism and subgraph isomorphism. In this paper we are introducing a new approach called SubGraD, for query graph detection in source graph. Firstly consider the model graph (query graph) and make the possible sets called model sets starting from the chosen initial node or starter. Similarly, for the source graph (reference graph), all the possible sets called reference sets could be made. Our aim is to make the reference set on the basis of the model set. If it is possible to make the reference set, then it is said that query graph has been detected in the source graph.*

## KEYWORDS

*model graph, reference graph, starter, model set, reference set*

## 1. INTRODUCTION

Graphs are data structures which are proved to be effective way of representing objects [1]. One of the problems of interest, with graphs, is matching a model graph (query graph) against a reference graph (source graph). Graph matching is the process of finding a correspondence between the nodes and the edges of two graphs that satisfies some (more or less stringent) constraints ensuring that similar substructures in one graph are mapped to similar substructures in the other [2]. Many graph databases are growing rapidly in size. The growth is both in the number of graphs and the sizes of graphs (the number of nodes and the number of edges). There is a critical need for efficient and effective graph querying tools for querying and mining these growing graph databases.

We are introducing a new method called SubGraD, which is helpful in graph (directed graph) mining. Here we take the model graph first and make the model sets. Suppose that there are two nodes in the model graph (query graph), and then corresponding model set will have only two elements. Then we break the reference graph in such a way that each set of reference set will have sets of two elements. We are assuming that the smallest model set is the set which has two elements corresponding to two nodes of the query graph. We are not considering the case where the self loop is present. In this paper, we are trying to make sets for reference set by using the SubGraD algorithm. If reference set could be made, then query graph is present in reference graph otherwise not. Related works are discussed in section 2. In section 3, SubGraD algorithm is given. Model sets have been made in section 4. And in section 5, it is tried to make Reference sets so that we could find out whether the query graph is present on the source graph or not.

## 2. RELATED WORK

Graph matching is the concept that has been intensively used in various applications such as in the field of social networks, road networks, network topology, chemical structures, graph grammars and semantic networks or pattern recognition. Some of the related works associated with different fields for graph mining is discussed in this section.

In Fuzzy set theory, there are so many works proposed for graph matching, for example Perchant and I. Bloch [3-6] gave methods for graph matching using fuzzy set theory , and Hwan gave an idea for sub-graph matching [7]. Fuzzy attributed graph models are suggested for very different image type representations, such as fingerprint verification [8].

Fernandez and Valiente proposed a method to represent attributed relational graphs, the maximum common subgraph and the minimum common super graph of two graphs by means of simple structures [9]. Bunke [10] suggested the relationship between graph edit distance and the maximum common subgraph, graph edit distance computation is equivalent to solving the maximum common subgraph problem.

Cross [11] described an outline for performing relational graph matching using genetic algorithm. He showed that Bayesian consistency measure could be proficiently optimized using a hybrid genetic search process that includes a local search strategy using a hill-climbing step. This hill-climbing step accelerates convergence considerably. Cross extended this idea in [12] that is also a convergence analysis for the problem of attributed graph matching using genetic search.

Myers and Hancock gave an idea that in attributed graph matching problems usually there are more than one valid and satisfactory solution, and they proposed a method to obtain different solutions at the same time during the genetic search, using an appropriately modified genetic algorithm [13-14].

Hancock and Kittler [15-16] used an iterative approach called probabilistic relaxation, and considering binary relations and assuming a Gaussian error. Later a Bayesian perspective was considered for both unary and binary attributes by Christmas [17], Gold and Rangarajan [18], Wilson and Hancock [19-20]. Williams [21] presented a comparative study of various deterministic discrete search-strategies for graph-matching, which was based on the previous Bayesian consistency measure in [19-20] and Tabu search was proposed as a graph matching algorithm.

## 3. SUBGRAD ALGORITHM

Graph Matching techniques are important and very general form of pattern matching that finds realistic use in areas such as image processing, pattern recognition and computer vision, graph grammars, graph transformation, bio computing, search operation in chemical structural formulae database, etc. SubGraD Algorithm is one such graph matching method. In SubGraD method, an adjacency matrix [A] is drawn for a query graph (model graph). Element $A_{ij}$ of [A] is 1 if an edge is present from node i to node j else 0. Using [A], model set M is prepared. Model set M is a set, consisting all sets (i, j) as its element where $A_{ij}$ is 1. On the basis of model set we will try to make the sets for reference set.

**SubGraD**

**For** i: 1 to number of nodes (n) in query graph

  **For j**: 1 to number of nodes (n) in query graph

    Make the adjacency matrix [A] of n×n

 **END** FOR j

**END** FOR i

Make Model Set M, M has the elements of the type $(a_1, b_1), (a_1, b_2), \ldots (a_n, b_n)$ where $A_{ab}=1$

    **For** i: 1 to n

  **For j**: 1 to n

    M= (i, j), where $A_{ij}=1$

 **END** FOR j

**END** FOR i

**For** i: 1 to number of nodes (m) in source graph

 **For j**: 1 to number of nodes (m) in source graph

    Make the adjacency matrix [S] of m×m

 **END** FOR j

**END** FOR i

Try to make the reference set R corresponding to Model Set M

Try to break adjacency matrix [S] into k sets $R_1, R_2, \ldots, R_k$, where $S_{ij}=1$ (row i and column j has entry 1) and each $R_k$ has the same number of elements as M has

R= ( $R_1, R_2, \ldots, R_k$)

If reference R set can be made, then query graph is detected in source graph

Otherwise query graph doesn't exist

**END SubGraD**

## 4. MODEL SETS

Suppose we have the query graphs as shown in Figure 1, and our purpose is to detect these query graphs in a given source graph. First step will be to draw the adjacency matrices for query graphs. In first query graph (Figure 1(a)), there are two nodes a and b, hence Adjacency matrix [A] will be 2×2. A single edge is present from a to b so the adjacency matrix [A] will have $A_{ab}$ entry as 1 and rest of the elements ($A_{aa}$, $A_{ba}$, $A_{bb}$) as 0. Likewise, the adjacency matrices Figure 2(b), 2(c) and 2(d) are drawn for query graphs Figure 1(b), 1(c) and 1(d) respectively.

Next step is to prepare Model sets using these adjacency matrices. The adjacency matrix of Figure 1(a) has four elements $A_{aa}$, $A_{ab}$, $A_{ba}$, $A_{bb}$ (Figure 2(a)). The model set corresponding to this adjacency matrix is (a, b), since only $A_{ab}$ entry is 1. Similarly for query graph (Figure 1(b)), the model set is (a, b) and (b, c), since there are two entries which are 1 in adjacency matrix (i.e. Figure 2(b)). Model sets for query graphs of Figure 1, is shown in TABLE 1.

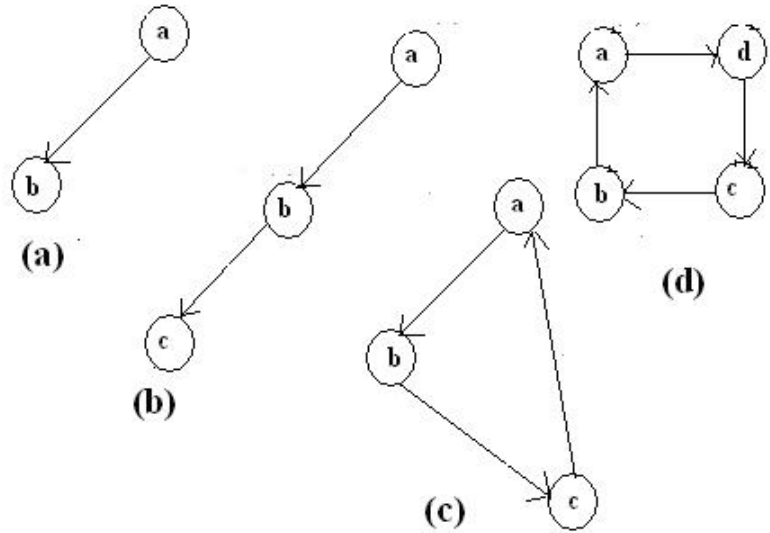

Figure1. Query graphs

Figure 2. Adjacency matrices for query graphs

Table 1. Model Sets for Query Graphs.

| Model set for Query Graph Figure 1(a) | Model set for Query Graph Figure1(b) | Model set for Query Graph Figure 1(c) | Model set for Query Graph Figure 1(d) |
|---|---|---|---|
| (a, b) | (a, b) (b, c) | (a, b) (b, c)(c, a) | (a, d) (b, a) (c, b) (d, c) Or (a, d) (d, c) (c, b), (b, a) |
| MODEL SET I | MODEL SET II | MODEL SET III | MODEL SET IV |

## 5. REFERENCE SETS

Let us consider the source graph present in Figure 3, now we will try to make reference set for source graph. Firstly select the Model set, we are taking MODEL SET I (from Table 1), there is only an edge that is present from node a to node b, so we are applying SubGraD algorithm, which selects the set of row i and column j, where the entry is 1 in adjacency matrix [S], of source graph (i.e. Figure 4). That means we are trying to make reference set R which consists of set $R_1, R_2, ..., R_n$. In source graph for row 1, and for column 2, 3, 4 and 6 the entries are 1, so the reference set for row 1 is (1, 2),(1, 3), (1, 4) (1, 6). Similarly for row 2, the reference set is (2, 4), (2, 6), for row 3, the reference set is (3, 5), for row 4, the reference set is (4, 6), for row 5, the reference set is (5, 1) and for row 6, the reference set is (6, 5). Corresponding to MODEL SET I, the Reference set R I (Table 2) have been found. Hence the query graph is detected in source graph.

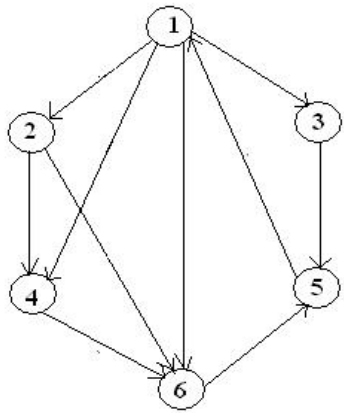

Figure 3. Source graph

|   | 1 | 2 | 3 | 4 | 5 | 6 |
|---|---|---|---|---|---|---|
| 1 | 0 | 1 | 1 | 1 | 0 | 1 |
| 2 | 0 | 0 | 0 | 1 | 0 | 1 |
| 3 | 0 | 0 | 0 | 0 | 1 | 0 |
| 4 | 0 | 0 | 0 | 0 | 0 | 1 |
| 5 | 1 | 0 | 0 | 0 | 0 | 0 |
| 6 | 0 | 0 | 0 | 0 | 1 | 0 |

Figure 4. Adjacency matrix of Source graph

Now suppose we want to find out whether query graph (Figure 1 (b)) is present or not, take the MODEL SET II from TABLE 1, i.e. (a, b), (b, c). In MODEL SET II we can see that element b is presented twice in the set. This type of element is called middle element. Actually, it is the node which has an incoming edge as well as an outgoing edge (in our case the node b has an incoming edge from node a and an outgoing edge to node c from Figure 1(b)). Now from source graph we have to make such a reference set $R = (R_1, R_2)$, where $R_1$ should have the set (i, k) and $R_2$ should have set (k, j). Here element k is the middle element. Firstly we start from row 1 in source graph (Figure 4), $S_{12}$, $S_{13}$, $S_{14}$ and $S_{16}$ are equal to 1, that means entries from row 1 to column 2, 3, 4 and 6 is 1. So next we will take element 2, 3, 4 and 6 as middle element, and will now consider row 2, row 3, row 4 and row 6. From adjacency matrix [S] (Figure 4), we will see that which Sij entry is 1, here i is equal to 2, 3, 4 and 6. For row 2, $S_{24}$ and $S_{26}$ elements are 1. So in consequence there are two elements, i.e. $S_{12}$ and $S_{24}$, which are 1. So the reference set R is made which has an element (1, 2), (2, 4), where 2 is the middle element. Similarly other elements of the reference set can be made. So reference set for row 1 is (1, 2)(2,4), (1, 2)(2, 6), (1, 3)(3, 5), (1, 4)(4, 6), (1, 6)(6, 5). For row 2, the reference set is (2, 4) (4, 6), (2, 6) (6, 5). For row 3, the reference set is (3, 5) (5, 1). For row 4, the reference set is (4, 6) (6, 5). For row 5, the reference set is (5, 1)(1, 2), (5, 1)(1, 3), (5, 1)(1, 4), (5, 1)(1, 6). For row 6, the reference set is (6, 5) (5, 1). For row 1, 2, 3, 4, 5 and 6 we can make the reference set for MODEL SET II, hence the query graph Figure 1(b) exists in the source graph at all nodes. Reference set II for MODEL SET II can be seen by Table 2.

Now suppose we want to find out whether query graph (Figure 1 (c)) is present or not, take the MODEL SET III from Table 1, i.e. (a, b) (b, c)(c, a). In MODEL SET III, we can see that there are middle elements b and c, and a cycle is formed. Cycle formation means that the starting node is also the ending node. It should be noted that the order of the set is not important, it means that, (a, b) (b, c) (c, a) is same as (a, b) (c, a) (b, c). But the former one satisfies the definition of cycle formation, so we should try to write in the order as the former one has. From adjacency matrix $S_{ij}$, we could see that, $S_{13}$, $S_{35}$, $S_{51}$, $S_{16}$, $S_{65}$, $S_{51}$ entries are 1. Reference set for row 1, is (1, 3)(3, 5)(5, 1) and (1, 6)(6, 5)(5, 1). Hence for node 1, the query graph has been detected twice. Elements 1, 3 and 5 corresponds to the elements a, b, c respectively, which forms a cycle. The null entry in Table 2 denotes that for that particular node, the query graph could not be detected. It could be seen from Table 2, for node 2 there is no reference set for MODEL SET III. Hence the query graph (Figure 1 (c)), is not present at node 2 of the source graph. Similarly for node 4 of source graph, this query graph cannot be detected.

For the query graph (Figure 1 (d)), again the cycle is formed. But here the model set M is consists of four sets. So now we will try to make reference set R which should have $R_1$, $R_2$, $R_3$ and $R_4$ in such a way that they should form a cycle. For example from adjacency matrix [S], it could be seen that $S_{46}$, $S_{65}$, $S_{51}$ and $S_{14}$ are 1, and they form a cycle. Hence query graph is detected at node 4. But at node 2 and node 3 it is not present. Hence if we could not able to make the reference set for particular query graph, then query graph could not be detected in the source graph.

Table 2. Reference sets for source graph.

| Nodes | Reference Set I for MODEL SET I | Reference Set II for MODEL SET II | Reference Set III for MODEL SET III | Reference Set IV for MODEL SET IV |
|---|---|---|---|---|
| 1 | i.(1, 2),<br>ii.(1, 3),<br>iii.(1,4),<br>iv.(1,6) | i.(1, 2)(2, 4)<br>(1, 2)(2, 6)<br>ii.(1, 3)(3, 5)<br>iii.(1, 4)(4, 6)<br>iv.(1, 6)(6, 5) | i.(1,3)(3,5)(5,1)<br>ii.(1,6)(6,5)(5,1) | i.(1,2)(2,6)(6,5)(5,1)<br>ii.(1,4)(4,6)(6,5)(5,1) |
| 2 | i.(2, 4),<br>ii.(2, 6) | i.(2,4)(4,6)<br>ii.(2,6)(6,5) | - | - |
| 3 | (3,5) | (3,5)(5,1) | (3,5)(5,1)(1,3) | - |
| 4 | (4,6) | (4,6)(6,5) | - | (4,6)(6,5)(5,1)(1,4) |
| 5 | (5,1) | i.(5,1)(1,2)<br>ii.(5,1)(1,3)<br>iii.(5,1)(1,4)<br>iv.(5,1)(1,6) | i.(5,1)(1,3)(3,5)<br>ii.(5,1)(1,6)(6,5) | (5,1)(1,2)(2,6)(6,5)<br>same as<br>(1,2)(2,6)(6,5)(5,1) |
| 6 | (6,5) | (6,5)(5,1) | (6,5)(5,1)(1,6)<br>same as<br>(1,6)(6,5)(5,1) | (6,5)(5,1)(1,2)(2,6)<br>same as<br>(1,2)(2,6)(6,5)(5,1) |

.

## 5. CONCLUSION

In this paper we have introduced an efficient approach for graph matching SubGraD. We first make the model set M for query graphs with the help of adjacency matrix and then corresponding to that model set, we try to make reference set R for source graph. If reference set could be made, then query graph is detected in source graph otherwise it does not exist in the source graph.

## ACKNOWLEDGEMENTS

I am very much thankful to Human Resource Development Group, CSIR for providing me a research internship to carry out this work.

## REFERENCES

[1]  A. Eshera and K.-S. Fu, (1986) "An image understanding using attributed symbolic representation and inexact graph matching", IEEE *Transactions on Pattern Analysis and Machine Intelligence*, 8(5):604–618.

[2]  D Conte, P Foggia, C Sansone, M Vento (2004) "Thirty years of graph matching in pattern recognition", *Int. Journal of Pattern Recognition and Artificial Intelligence* **18,** 265–298.

[3]  Perchant and I. Bloch. (1999) "A new definition for fuzzy attributed graph homomorphism with application to structural shape recognition in brain imaging", In IMTC'99, *16th IEEE Instrumentation and Measurement Technology Conference*, pages 1801–1806, Venice, Italy.

[4]  Perchant and I. Bloch. (2000) "Graph fuzzy homomorphism interpreted as fuzzy association graphs", In Proceedings of the *International Conference on Pattern Recognition*, ICPR 2000, volume2, pages 1046-1049, Barcelona, Spain.

[5]  Perchant and I. Bloch. (2000) "Semantic spatial fuzzy attribute design for graph modelling", In $8^{th}$ *International Conference on Information Processing and Management of Uncertainty in Knowledge based Systems* IPMU 2000, volume 3, pages 1397–1404, Madrid, Spain.

[6]  Perchant and I. Bloch. (2002) "Fuzzy morphisms between graphs", *Fuzzy Sets and Systems*, 128 (2) pages 149–168.

[7]  Sung Hwan. (2001) "Content-based image retrieval using fuzzy multiple attribute relational graph", *IEEE International Symposium on Industrial Electronics Proceedings* (ISIE 2001), 3:1508–1513.

[8]  Fan, C. Liu, and Y. Wang. (2000) "A randomized approach with geometric constraints to fingerprint verification", *Pattern Recognition*, 33(11):1793–1803.

[9]  Fernandez and G. Valiente. (2001) "A graph distance metric combining maximum common subgraph and minimum common supergraph", *Pattern Recognition Letters*, 22(6-7):753– 758.

[10]  Bunke. (1997) "On a relation between graph edit distance and maximum common subgraph" *Pattern Recognition Letters,* 18(8):689–694.

[11]  D. J. Cross, R. C. Wilson, and E. R. Hancock. (1997) "Inexact graph matching using genetic search", *Pattern Recognition*, 30(6):953–970.

[12]  D. J. Cross, R.Myers, and E. R. Hancock. (2000) "Convergence of a hill-climbing genetic algorithm for graph matching", *Pattern Recognition*, 33(11):1863–1880.

[13] R. Myers and E. R. Hancock. (2000) "Genetic algorithms for ambiguous labelling problems", *Pattern Recognition*, 33(4):685–704.

[14] R. Myers and E. R. Hancock. (2001) "Least-commitment graph matching with genetic algorithms", *Pattern Recognition*, 34(2):375–394.

[15] R. Hancock and J. Kittler. (1990) "Edge-labeling using dictionary-based relaxation", *IEEE Transactions on Pattern Analysis and Machine Intelligence*, 12(2):165–181.

[16] Kittler, W. J. Christmas, and M. Petrou. (1993) "Probabilistic relaxation for matching problems in computer vision", *IEEE Proceedings of the International Conference on Computer Vision* (ICCV93), pages 666–673.

[17] J. Christmas, J. Kittler, and M. Petrou. (1995) "Structural matching in computer vision using probabilistic relaxation" *IEEE Transactions on Pattern Analysis and Machine Intelligence*, 17(8):749–64.

[18] Gold and A. Rangarajan. (1996) "A graduated assignment algorithm for graph matching", *IEEE Transactions on Pattern Analysis and Machine Intelligence*, 18(4):377–88.

[19] C. Wilson and E. R. Hancock. (1996) "Bayesian compatibility model for grach matching", *Pattern Recognition Letters*, 17:263–276.

[20] R. C. Wilson and E. R. Hancock. (1997) "Structural matching by discrete relaxation", *IEEE Transactions on Pattern Analysis and Machine Intelligence*, 19(6):634–698.

[21] Williams, R. C. Wilson, and E. R. Hancock. (1999) "Deterministic search for relational graph matching", *Pattern Recognition*, 32(7):1255–1271.

**Authors**

Ms. Akshara Pande is a Research Intern in HRDG, CSIR. She has done M.Sc. in Computer Science from J.K. Institute of Applied Physics & Technology, University of Allahabad, Allahabad. Her research area is Design Pattern Detection and Subgraph Detection.

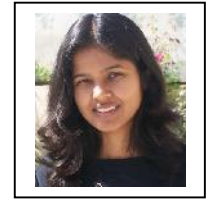

Mr. Vivekanand Pant is a Senior Software Engineer in IBM Noida. He has eight years industry experience. He has BTech in Information Technology from Motilal Nehru National Institute of Technology, Allahabad. He is currently working in software development area.

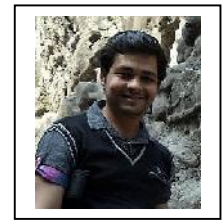

Mr. Shailendra Nigam holds a Masters degree in Electronics Science and is currently working as a Sr. Principal Scientist in HRDG, CSIR. His research interests include DSP, Text to Speech

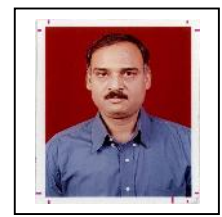

Synthesis, PC Based System Design and Workflow Automation. He holds a patent and has filed many software copyrights.